\documentclass[11pt]{article}

\usepackage{natbib,hyperref,ragged2e,graphicx}

\title{CybergeoNetworks, an interactive application for the geographical and semantic analysis of scientific publications}
\author{C. Cottineau$^{1,2}$, J. Raimbault$^{3,4,1,\ast}$, P.-O. Chasset$^{5}$, H. Commenges$^{1}$,\\
A. Banos$^{6}$ and D. Pumain$^{1}$\medskip\\
$^{1}$ UMR CNRS 8504 G{\'e}ographie-cit{\'e}s, France\\
$^{2}$ UMR CNRS 8097 Centre Maurice Halbwachs, France\\
$^{3}$ CASA, UCL, United Kingdom\\
$^{4}$ UPS CNRS 3611 ISC-PIF, France\\
$^{5}$ LISER, Luxembourg\\
$^{6}$ UMR IDEES 6266, France\medskip\\
$\ast$ \texttt{juste.raimbault@polytechnique.edu}
}

\date{}

\begin{document}

{\justify \textit{Translated from: }Cl{\'e}mentine Cottineau, Juste Raimbault, Pierre-Olivier Chasset, Hadrien Commenges, Arnaud Banos, and Denise Pumain (2017). CybergeoNetworks, une application interactive pour l'analyse g{\'e}ographique et s{\'e}mantique des publications scientifiques. In Bouzeghoub M., Mosseri R. (eds.) \textit{Les Big Data {\`a} d{\'e}couvert}, CNRS Editions (EAN: 9782271114648), pp.272-273.}

{\let\newpage\relax\maketitle}

\begin{abstract}
	The increase in the number of publications has made more difficult for authors to situate their work within previous literature, especially on subjects studied from different disciplinary viewpoints. Besides, new data analysis techniques and new bibliometrics data sources provide an opportunity to map and navigate scientific landscapes. We introduce here an open-source and open-access web application designed for the multi-dimensional exploration of a journal content, including the mapping of geographical, semantic and citations networks. The application is profiled and implemented for the geography journal Cybergeo, a generalist geography journal which receives contributions from multiple sub-disciplines. We suggest that such initiatives are crucial to promote open science and reflexivity.
\end{abstract}

\bigskip

The exponential growth in the number of published papers and the increasing number of journals have led scientific publication into an era of big data. The evolutions induced by the use of new information technologies seem more to have contributed to this massive growth of references rather than having simplified editorial processes or facilitated user access to the scientific literature. The difficulties in dealing with these data go far beyond what Eugene Garfield had anticipated when creating in 964 the ISI database (Institute for Scientific Information, which later became Web of Science after being bought by Thomson Reuters) \citep{garfield1970citation}. Progressively, the benefits of ``peer'' work which is done as volunteers to evaluate the scientific production before publication, mostly by providing a contextualisation into existing knowledge, has been captured by publishers. These were supposed to guarantee quality but punctured academic libraries and built barriers to knowledge diffusion, despite the public calls for an open access to science.

\bigskip

The issue of mastering existing literature remains crucial for any scientific who aims at knowing and surveying a ``state-of-the-art''. It becomes even more difficult for subjects situated at the interface of several disciplines, and which risk, if habitual disciplinary ``niches'' are followed, to be treated in an incomplete way. This may undermine the reaching potential of solutions that science can bring to societal problems. Geographers are particularly aware of such pitfalls since they have for long built a discipline at the boundaries of natural and social sciences, on settlements and urbanism issues, on the environment and health, on planning and development \citep{kosmopoulos2007citation}. This may not be a coincidence if a tool proposed to help improving the exploration of a publication universe is elaborated from a geography journal, Cybergeo.

\bigskip

The progresses made in terms of data collection and analysis of big data make it possible today to navigate networks of scientific publications in a novel way. We designed and implemented an original open-access application, which enables the exploration of text contents and keywords in more than 900 published papers for 20 years by Cybergeo. It also includes references cited in these papers, papers citing these or which cite the same references, among all other scientific journals accessible in Google Scholar, what corresponds to a database of around 200,000 papers with their hundreds of thousands associated keywords. The objective is to allow web users to realise themselves on demand maps of geographical and semantic proximities between publications and between geographical areas by exploring around this corpus.

\bigskip

The application is available at \url{https://analytics.huma-num.fr/geographie-cites/cybergeonetworks/}.

\bigskip

A first possibility offered by the application is to realise a diachronic mapping which represents for a given period the affiliation countries for authors and the countries studied in the paper. When this information is crossed with the thematic profile of papers, this reveals a diversity of research interests depending on localisation, and a semantic proximity between countries studied with similar terms. The spatial proximity and the semantic proximity are in some places coinciding. Thus for example, institutional Europe is identified as a space of string semantic proximity in which the lexicon of boundaries is over-represented and the lexicon of risk is under-represented (Figure 1).

\begin{figure}[h!]
	\includegraphics[width=\linewidth]{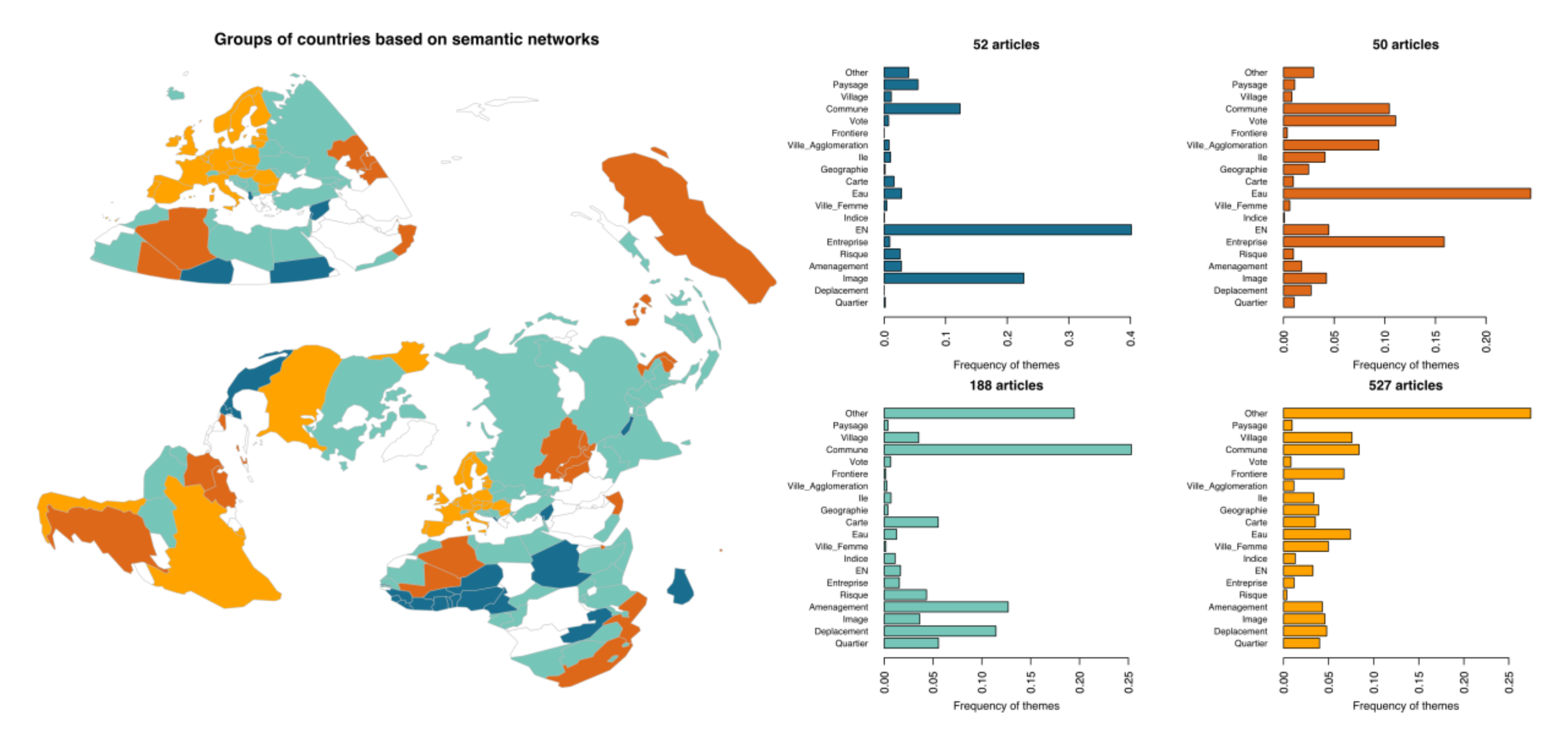}	
	\caption{Classification of countries studies in Cybergeo, as a function of the thematic profile of terms used in main text.}
\end{figure}

\bigskip

Keywords of papers in the journal and of cited papers allow connecting the articles using them and to build weighted networks depending on the number of these co-occurences \citep{chavalarias2013phylomemetic}. The structure of these networks provides a fine information on proximities between thematics, gathered into ``communities''. Such communities are identified by colours in Figure 2, and can be explored at different levels of granularity by zooming of the graph of the network in the application.

\begin{figure}[h!]
	\includegraphics[width=\linewidth]{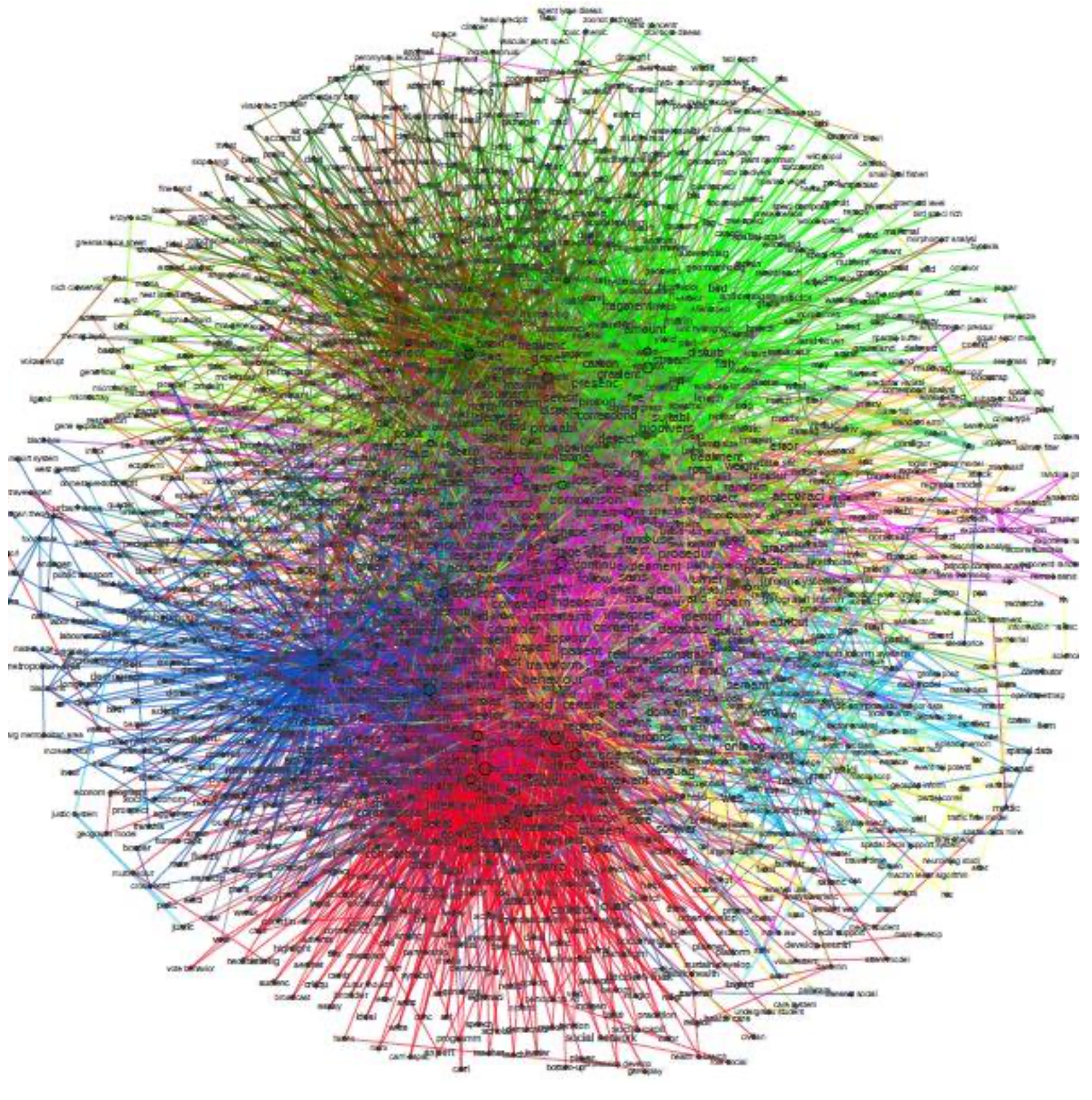}	
	\caption{Keywords network within papers cited in Cybergeo or citing them.}
\end{figure}

\bigskip

The Figure 2 spatialises this network and shows that disciplines such as physical geography and economic geography are linked through their common practice of methods such as spatial analysis and statistics or complexity paradigms.

\bigskip

\begin{figure}[h!]
	\includegraphics[width=\linewidth]{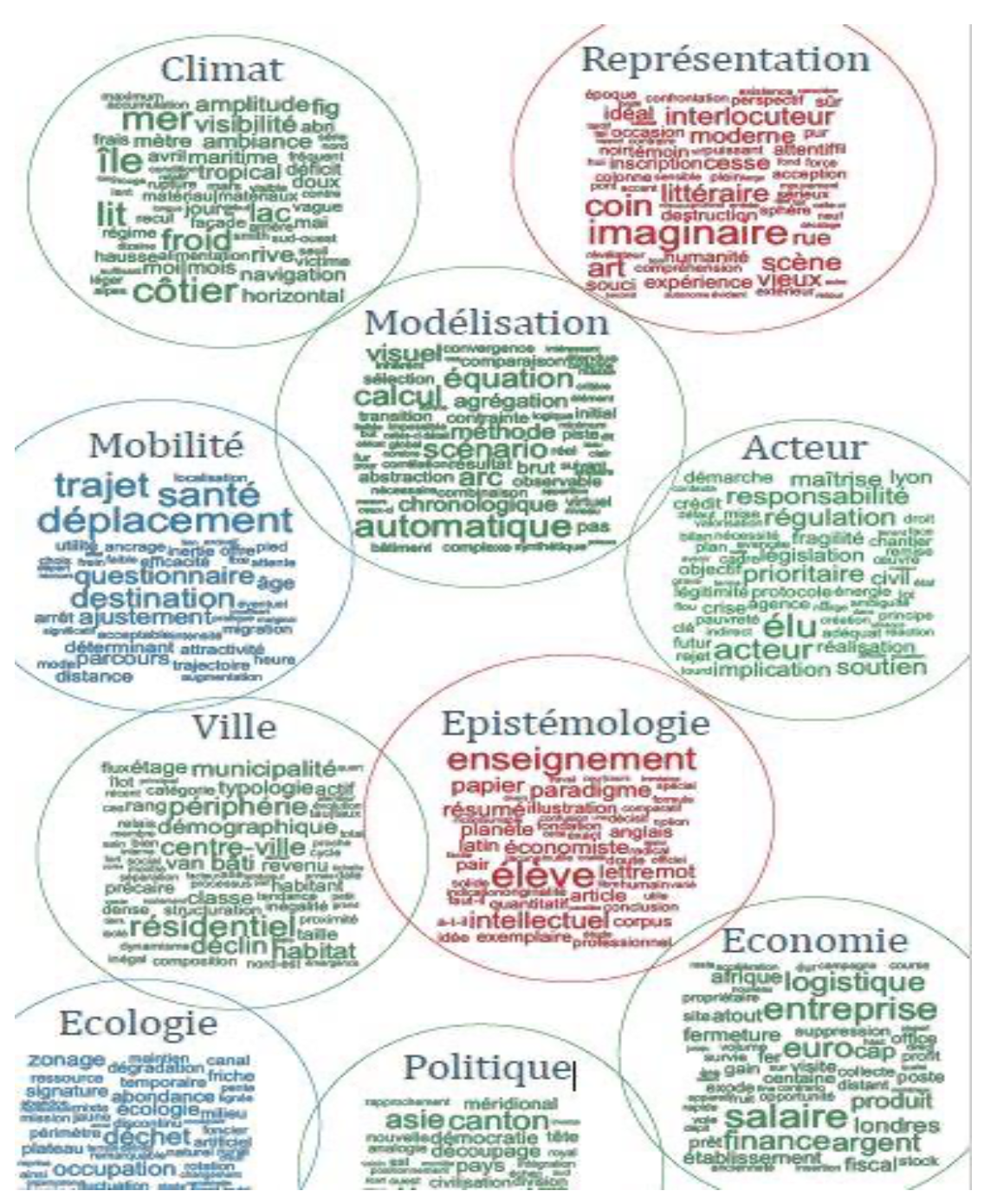}	
	\caption{Around ten broad themes obtained through the analysis of Cybergeo full texts and gathered into word clouds. The colour varies depending on the number of documents within each theme; the size of words in word clouds is proportional to the frequency of the word in the cloud theme.}
\end{figure}

The semantic analysis of paper full texts also allows constructing word clouds which gathers in synthetic themes. The granularity of theme grouping can be varied and the frequency of words within a theme can be measured (Figure 3).

\bigskip

The utility of this tool for authors is to improve their knowledge of themes studied in the journal, and to situate the potential contribution of a new paper in the universe of its references and of neighbour disciplines which can be influenced. For the editorial team, it provides an instrument to manage the journal policy, which can choose to maintain a rather generalist editorial line or to become more specialised. Finally, the application can be adapted to other online journals.

\bigskip

The large scientific publishing companies propose bibliometrics analysis tools, allowing researchers and journals to optimise their investment to be the best positioned on the citation market. We propose here an open and free tool for self-analysis, conceived to foster the reflexivity and integrity of research.

\bigskip


\end{document}